\documentclass[prl, lettersize,twocolumn,superscriptaddress,preprintnumbers,nofootinbib,showpacs]{revtex4-1}
\usepackage{graphicx}
\usepackage{amsmath}
\usepackage{amssymb}
\usepackage{color}
\usepackage{hyperref}
\usepackage{url}
\usepackage{breakurl}
\usepackage{float}
\usepackage{epstopdf}
\usepackage{extarrows}
\usepackage[normalem]{ulem}

\newcommand{\beq}{\begin{equation}}
\newcommand{\eeq}{\end{equation}}
\newcommand{\be}{\begin{equation}}
\newcommand{\ee}{\end{equation}}
\newcommand{\bea}{\begin{eqnarray}}
\newcommand{\eea}{\end{eqnarray}}
\newcommand{\nn}{\nonumber}

\newcommand{\mhh}{m_{hh}}

\begin{document}

\title{Resolving the Degeneracy in Single Higgs Production with Higgs Pair Production}

\author{Qing-Hong Cao}
\email{qinghongcao@pku.edu.cn}
\affiliation{Department of Physics and State Key Laboratory 
of Nuclear Physics and Technology, Peking University, Beijing 100871, China}
\affiliation{Collaborative Innovation Center of Quantum Matter, Beijing, China}
\affiliation{Center for High Energy Physics, Peking University, Beijing 100871, China}

\author{Bin Yan}
\email{binyan@pku.edu.cn}
\affiliation{Department of Physics and State Key Laboratory of 
Nuclear Physics and Technology, Peking University, Beijing 100871, China}

\author{Dong-Ming Zhang}
\email{zhangdongming@pku.edu.cn}
\affiliation{Department of Physics and State Key Laboratory of 
Nuclear Physics and Technology, Peking University, Beijing 100871, China}

\author{Hao Zhang}
\email{zhanghao@physics.ucsb.edu}
\affiliation{Department of Physics, University of California, 
Santa Barbara, CA 93106, USA}

\begin{abstract}
The Higgs boson production can be affected by several anomalous 
couplings, e.g. $c_t$ and $c_g$ anomalous couplings. 
Precise measurement of $gg\to h$ production yields two 
degenerate parameter spaces of $c_t$ and $c_g$; one parameter 
space exhibits the SM limit while the other does not. Such a 
degeneracy could be resolved by Higgs boson pair production.  
In this work we adapt the strategy suggested by the 
ATLAS collaboration to explore the potential of distinguishing 
the degeneracy at the 14 TeV LHC.  If the $c_t$ anomalous coupling
is induced only by the operator $H^\dag H \bar 
Q_L \tilde{H} t_R$,  then the non-SM-like band could be excluded
with an integrated luminosity of $\sim 235~{\rm fb}^{-1}$.   
Making use of the fact that the Higgs boson pair is mainly produced 
through an $s$-wave scattering, we propose an analytical function 
to describe the fraction of signal events surviving a series 
of experimental cuts for a given invariant mass of Higgs boson pair. 
The function is model independent and can be applied to estimate 
the discovery potential of various NP models. 
\end{abstract}

\maketitle

\begin{figure}
\includegraphics[scale=0.42]{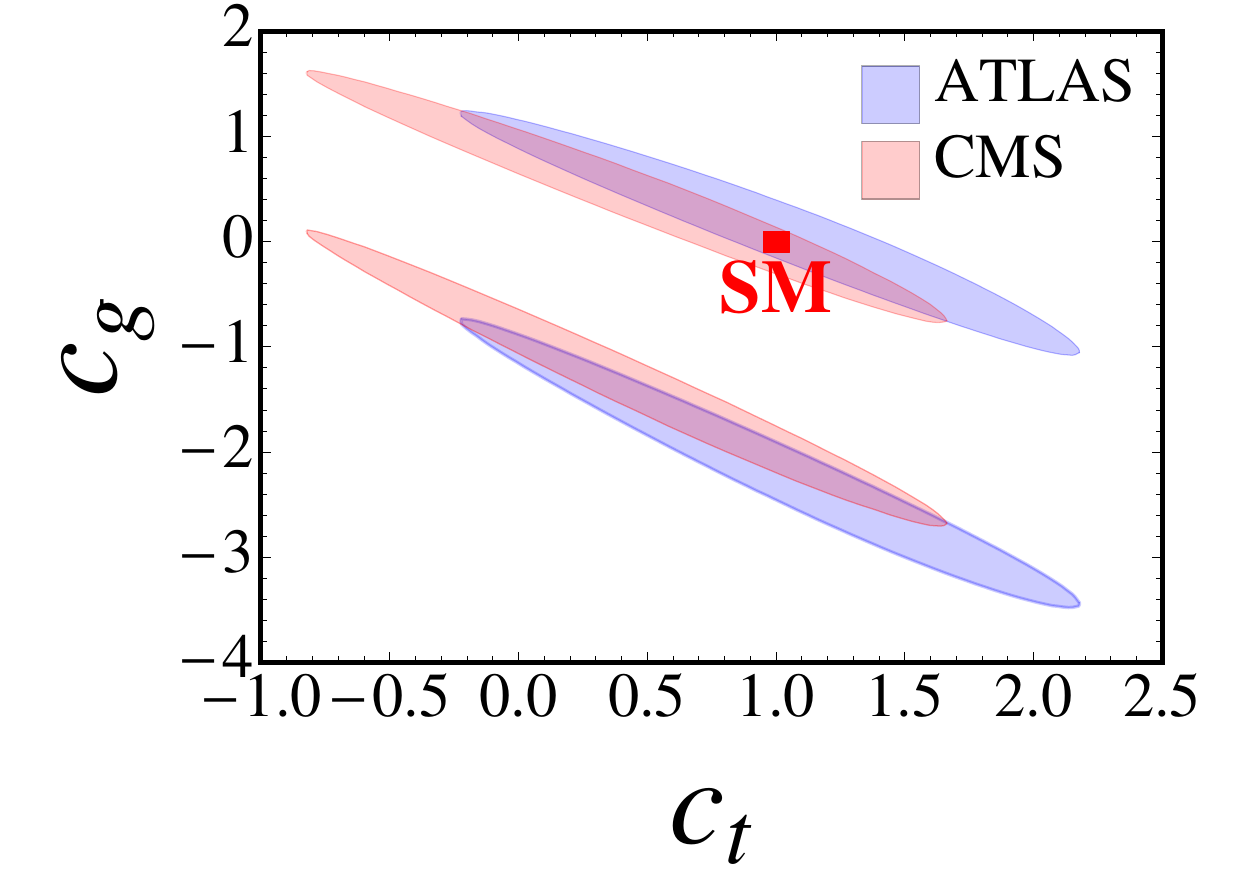}
\caption{The bounds at 95\% confidence level (C.L.) in the plane 
of $c_t$ and $c_g$ from the ATLAS~\cite{Aad:2015gba} and 
CMS~\cite{Khachatryan:2014jba} collaborations. The red box 
denotes the SM ($c_t=1, c_g=0$).}
\label{fig:globalfit}
\end{figure}

\noindent{\bf Introduction:~}%
In the establishment of the Standard Model (SM), thousands of experiments have been performed to measure model parameters and check the consistency of the theory. In the same spirit, it is critical to measure all the properties of the recently discovered Higgs boson as precisely as possible to test the SM and probe New Physics (NP) beyond the SM. 
The Higgs couplings have been constrained 
at the LHC Run-1~\cite{Aad:2015gba,Khachatryan:2014jba}. 
In the SM the Higgs boson is predominantly 
produced through the gluon fusion process 
which can be affected by either $ht\bar t$ or $hgg$ anomalous couplings. 
The anomalous couplings are described by the effective Lagrangian 
$\mathcal{L}= -\displaystyle{\tfrac{m_t}{v}} c_t h \bar{t}t  
+\displaystyle{\tfrac{\alpha_s}{12\pi v}}c_g h G^a_{\mu\nu}
G^{\mu\nu}_a$. 
Figure~\ref{fig:globalfit} displays the parameter space of 
$c_t$ and $c_g$ allowed by the measurement. 
Two degenerate parameter spaces arise from the interference of 
$c_t$ and $c_g$ contributions. For example, in the heavy top-quark limit, 
\beq
\sigma_{\rm NP}(gg\to h) \simeq \sigma_{\rm SM}(gg\to h)\times |c_g+c_t|^2.
\eeq
The upper band in Fig.~\ref{fig:globalfit} corresponds to $c_t +c_g 
\sim  +1$, which has a SM limit of $c_t \to 1$ and $c_g\to 0$. We name it as ``SM-like" band. 
The lower band corresponds to $c_t + c_g \sim -1$, which does not exhibit the SM limit. 
In the lower band the NP contributions should cancel the SM contribution out and 
leads to a residual contribution of minus SM value. 
Though often ignored, it is possible in principle
\footnote{
For example, additional colored $SU(2)_L$ singlet scalars 
($S_i$) could generate a large negative $c_g$ in the FNNP band. 
The scalar $S_i$'s interact with the SM Higgs boson via 
$-k_iS_i^*S_iH^\dag H$ with $H$ the SM Higgs boson doublet. 
Integrating out heavy $S_i$'s inside the $gg\to h$ triangle loop 
yields $c_g\sim \sum_{i}T(S_i)k_iv^2/(4m_{S_i}^2)$ where $T(S_i)$ 
is the Dynkin index of $S_i$ for the corresponding representation 
under $SU(3)_C$. }.
We name the lower band as the ``faked-no-new-physics'' 
(FNNP) band since further improvement in 
the measurement of single Higgs-boson productions cannot 
distinguish it from the upper SM-like band. 
The Higgs boson pair production $gg\to hh$, which is highly 
correlated with the $gg\to h$ process, can be used to 
discriminate the SM-like and FNNP bands~\cite{Chang:2013cia, Azatov:2015oxa,cao:2015xxx}. In this work, we focus our attentions on the ATLAS constraints and  
explore the potential of the 14~TeV LHC to exclude the FNNP band in 
the Higgs boson pair production. 

\noindent{\bf Higgs boson pair production:~}%
The Higgs boson pair production is usually considered as the best channel to measure the Higgs trilinear coupling in the SM~\cite{Glover:1987nx,Baur:2002rb,Baur:2002qd,Dolan:2012rv,Papaefstathiou:2012qe,Baglio:2012np,Goertz:2013kp,Barr:2013tda,Barger:2013jfa,Li:2013flc,deLima:2014dta,Slawinska:2014vpa,
Englert:2014uqa,Liu:2014rva,Barr:2014sga,Li:2015yia,Papaefstathiou:2015iba,He:2015spf}.
It is also sensitive to various NP models~\cite{Plehn:1996wb,Djouadi:1999rca,Belyaev:1999mx,Belyaev:1999vz,Belyaev:1999kk,Lafaye:2000ec,BarrientosBendezu:2001di,
Baur:2003gpa,Baur:2003gp,Liu:2004pv,Moretti:2004wa,Binoth:2006ym,deSandes:2007my,Wang:2007nf,
Ma:2009kj,Arhrib:2009hc,Han:2009zp,Asakawa:2010xj,Grober:2010yv,Figy:2011yu,
Gillioz:2012se,Kribs:2012kz,Dawson:2012mk,Dolan:2012ac,
Cao:2013si,Nhung:2013lpa,Ellwanger:2013ova,Nishiwaki:2013cma,Liu:2013woa,Haba:2013xla,Enkhbat:2013oba,
Kumar:2014bca,Xiao-ping:2014npa,Chen:2014xra,Chen:2014xwa,Ahriche:2014cpa,Barger:2014taa,Barger:2014qva,Cao:2014kya,
Dawson:2015oha,Enkhbat:2015bca,Etesami:2015caa}.
In this work we adapt the effective Lagrangian approach to describe the unknown NP effects. 
After the electroweak symmetry breaking the effective Lagrangian related to 
the double Higgs production is~\cite{Pierce:2006dh,Kanemura:2008ub, 
Contino:2012xk,Goertz:2014qta,Edelhaeuser:2015zra,
Azatov:2015oxa,Lu:2015jza}
\bea
\mathcal{L}_{h}&=&-\frac{m_h^2}{2v}c_3h^3
-\frac{m_t}{v}c_t\bar{t}_Lt_R h
-\frac{m_t}{v^2}c_{2h}\bar{t}_Lt_R h^2 \nn\\
&+&\frac{\alpha_s c_g}{12 \pi v}hG_{\mu\nu}^aG^{\mu\nu}_a
+\frac{\alpha_s c_g}{24 \pi v^2}h^2G_{\mu\nu}^aG^{\mu\nu}_a+{\rm h.c.}~,
\eea
where $v=246~{\rm GeV}$ is the vaccum expectation value, $\alpha_s=g_s^2/4\pi$ with $g_s$ the strong coupling strength, $m_t$ is the top-quark mass and $m_h$ is the Higgs boson mass. 
In the SM, $c_3=c_t=1$ and $c_{2h}=c_g=0$. 
The squared amplitude of $gg\to hh$ averaging over the gluon polarizations and colors is~\cite{Dawson:2015oha}
\bea
\overline{|{\cal M}|^2} = \frac{\alpha_s^2 \hat{s}^2}{256\pi^2 v^4}
\bigg[ && \bigg|  {3m_h^2\over {\hat s}-m_h^2} c_3
\left(c_t F_\bigtriangleup+{2\over 3} c_g\right) 
 +2c_{2h} F_\bigtriangleup \nn\\ 
&&+c_t^2 F_\Box+{2\over 3}c_g \bigg|^2+ \left|c_t^2  G_\Box\right|^2\bigg],
 \label{eq:hhdif}
\eea
where $F_\bigtriangleup\equiv F_\bigtriangleup({\hat s},{\hat t},m_h^2, m_t^2)$, 
$F_\Box\equiv F_\Box({\hat s}, {\hat t}, m_h^2, m_t^2)$ and 
$G_\Box\equiv G_\Box({\hat s},{\hat t}, m_h^2, m_t^2)$ are the 
form factors~\cite{Plehn:1996wb} with $\hat{s}$ and 
$\hat{t}$ the canonical Mandelstam 
variables. $G_\Box$ corresponds to the $d$-wave component 
which is negligible~\cite{Dawson:2012mk}.

\begin{figure}[b]
\includegraphics[scale=0.22,clip]{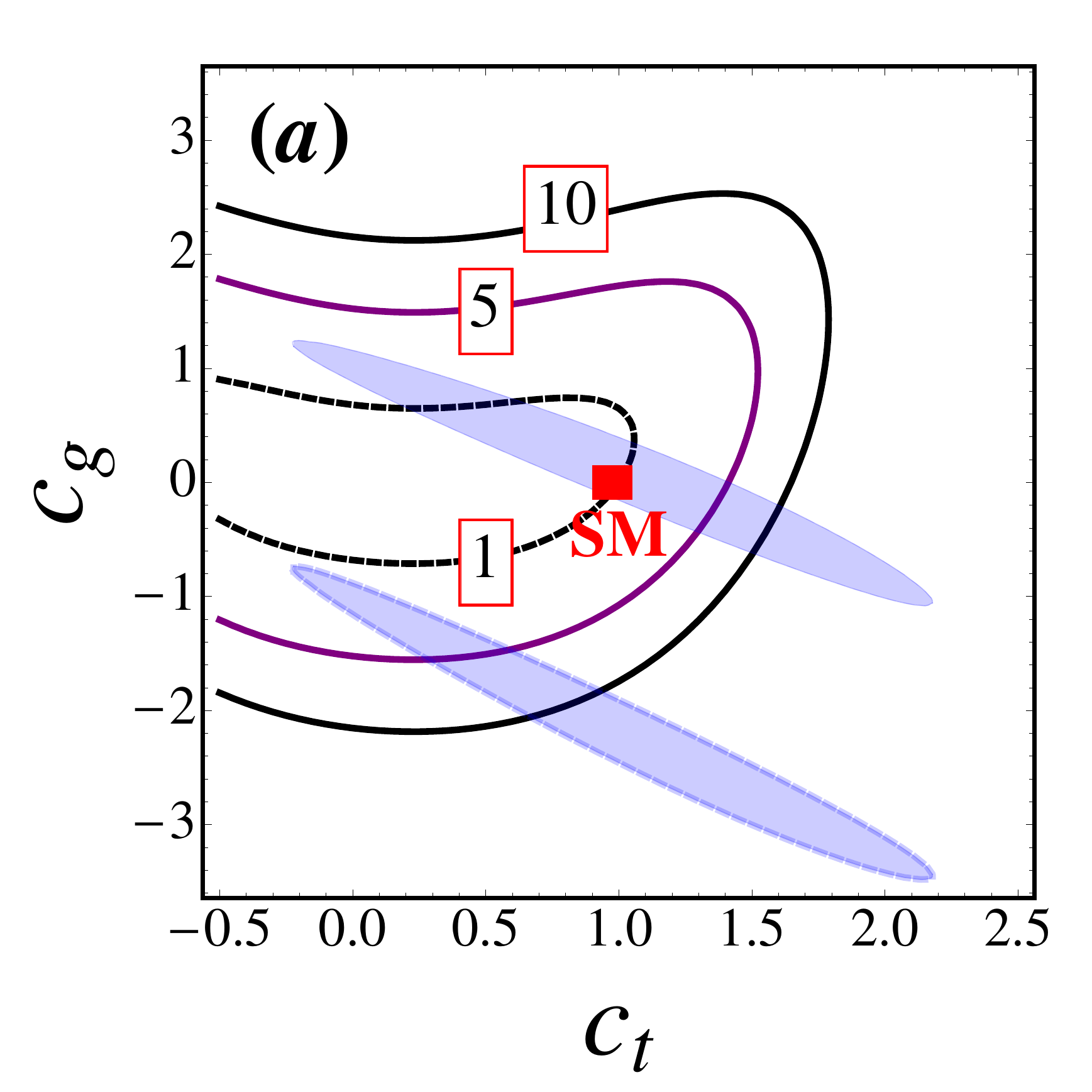}  
\includegraphics[scale=0.22,clip]{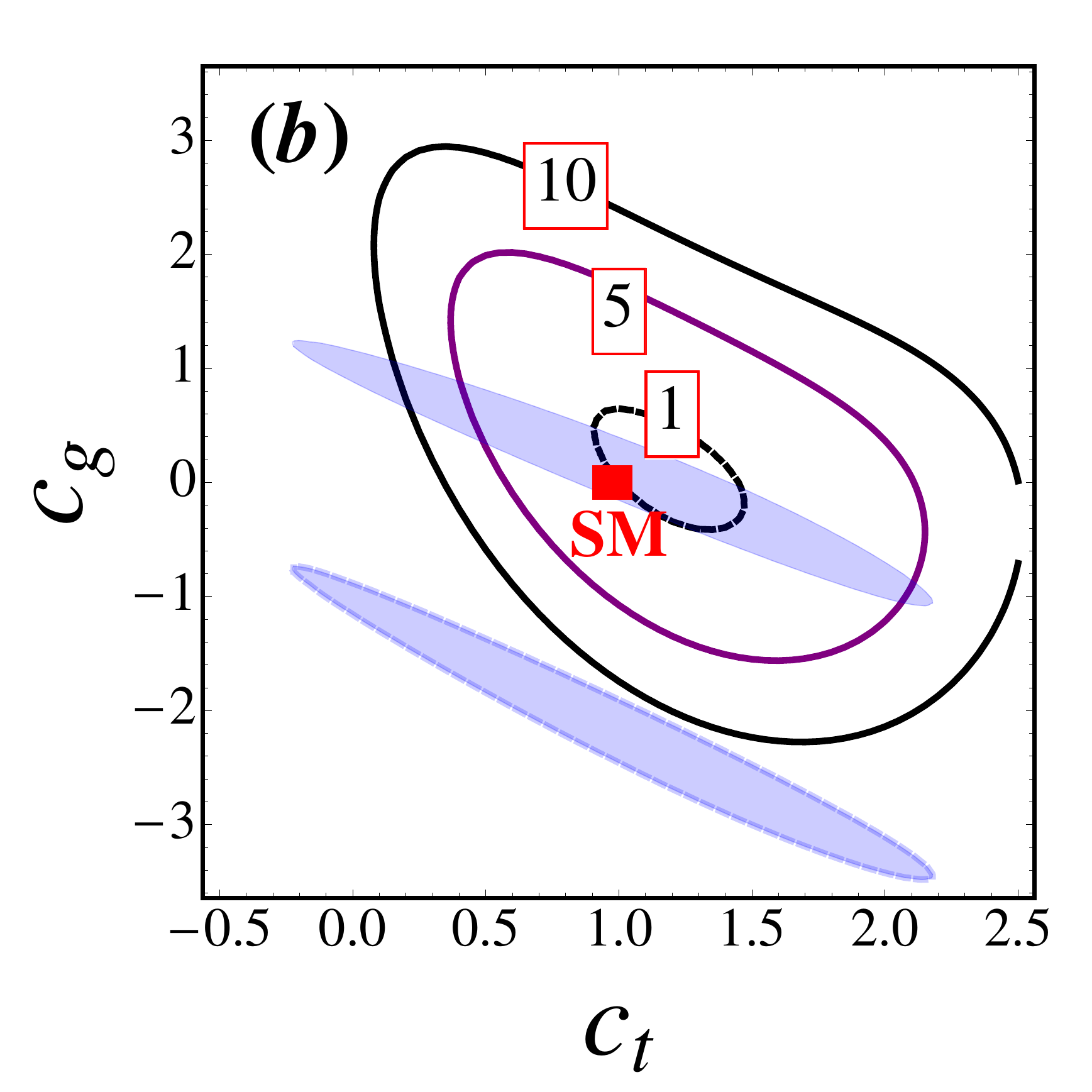}
\includegraphics[scale=0.22,clip]{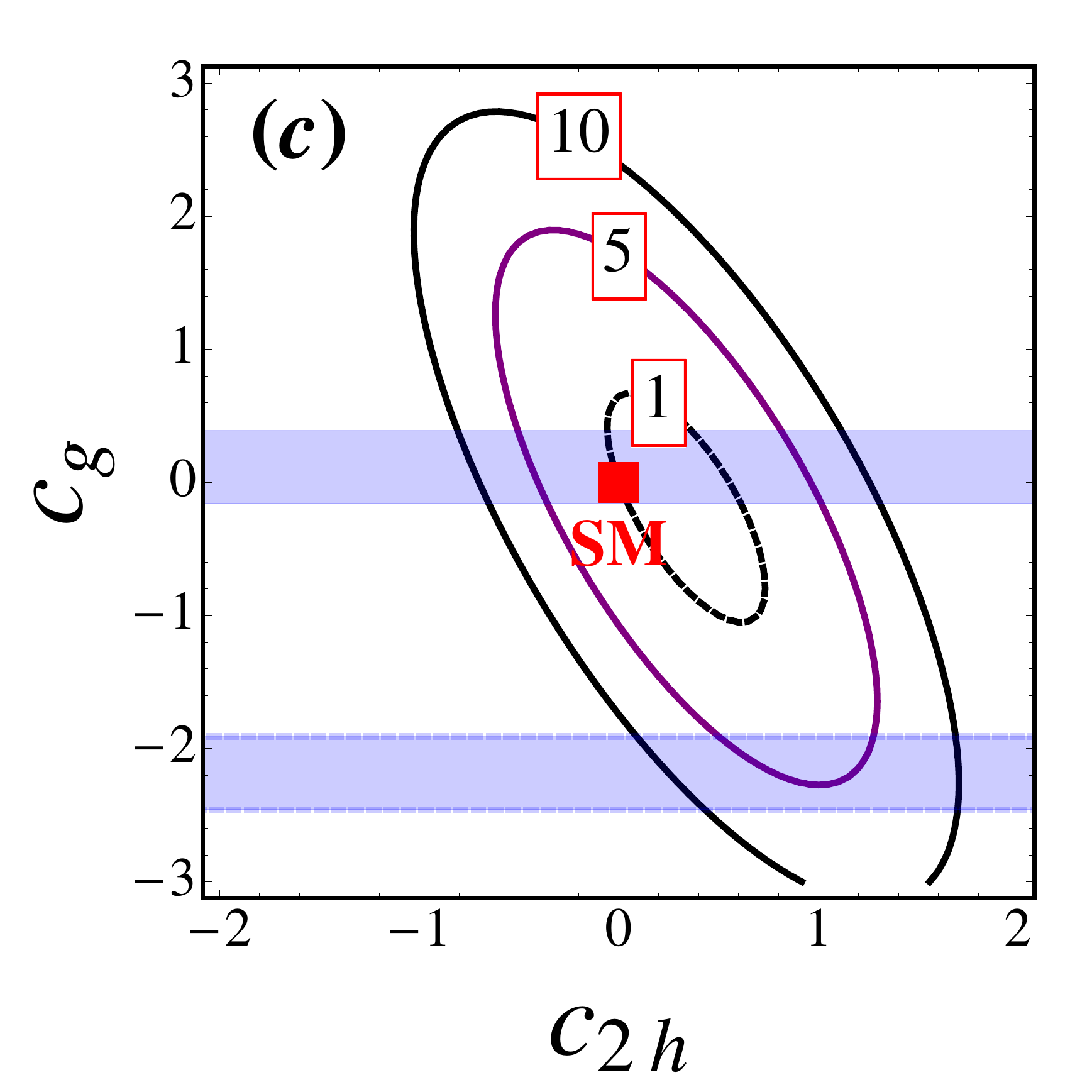}
\includegraphics[scale=0.22,clip]{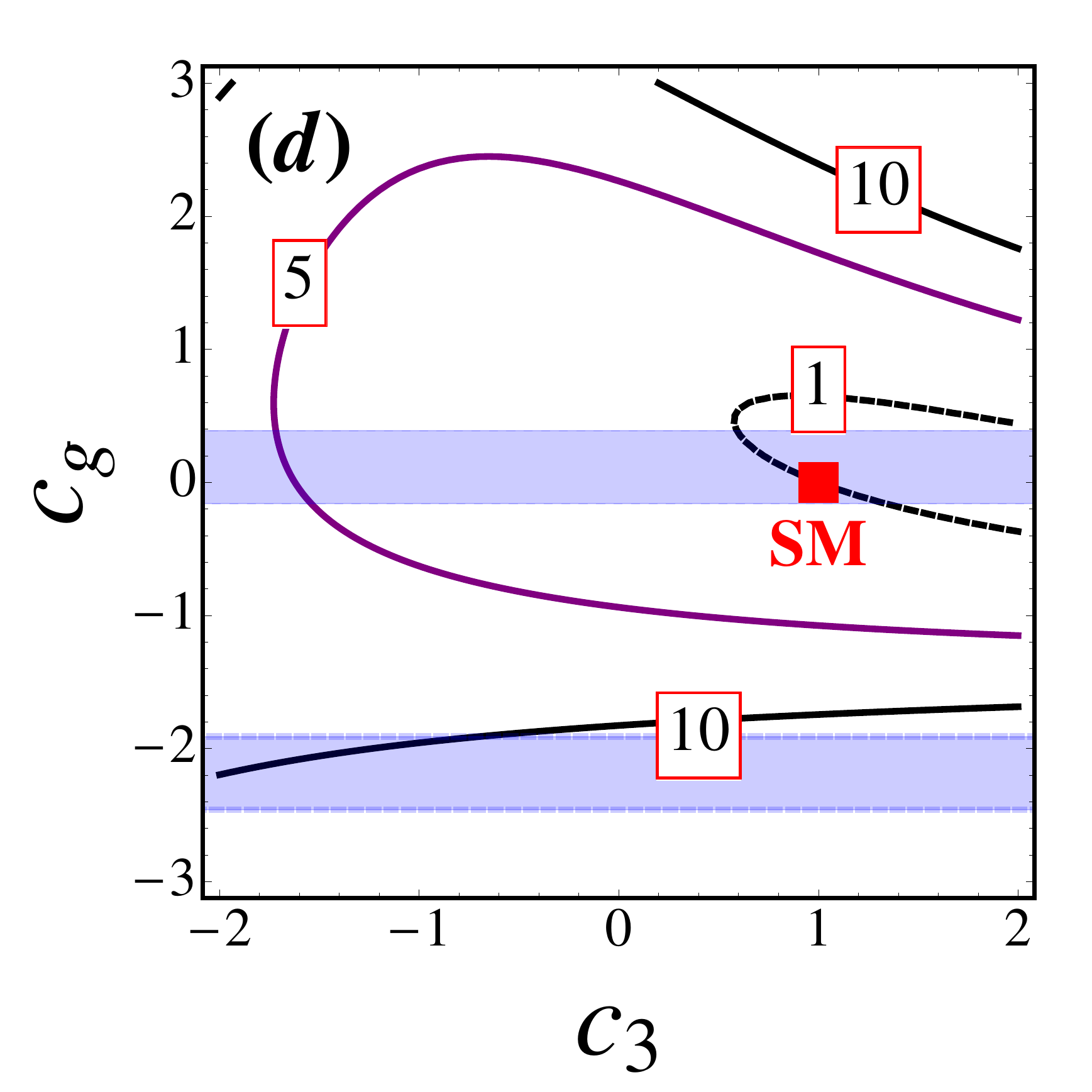}
\caption{The contours of $R_{hh}=1,~5$ and 10 in 
the plane of anomalous couplings at the 14 TeV LHC.
The figure indices correspond to the benchmark choices of anomalous couplings in Eq.~\ref{eq:benchmark}.
\label{fig:R14} }
\end{figure}

In order to compare $\sigma(gg\to hh)$ with the 
SM prediction, we define a ratio $R_{hh}$ as
\bea
R_{hh}&=&\frac{\sigma (gg\to hh)}{\sigma^{\rm SM} (gg\to hh)}.
\eea
Figure~\ref{fig:R14} displays the contours of $R_{hh}=1$, 5 and 10 
in the plane of anomalous couplings for four benchmark choices: 
\begin{align}
&(a) ~ c_3=1, c_{2h}=0; && (b) ~ c_3=1, c_{2h}=3(c_t-1)/2; \nn\\
&(c) ~ c_3=c_t=1; &&(d) ~ c_t=1,  c_{2h}=0. \label{eq:benchmark}
\end{align}
In the case $(b)$ we assume that both $c_t$ and $c_{2h}$ 
anomalous couplings are induced by the operator $H^\dag H \bar 
Q_L \tilde{H} t_R$ and thus exhibit the relation. 
The production cross section of the Higgs boson pair is 
enhanced in the FNNP band in all the four choices: 
\begin{align}
& (a) ~1.44<R_{hh}<92.89; && (b) ~11.12<R_{hh}<46.58; \nn\\
& (c) ~3.46<R_{hh}<88.86; && (d)  ~ 9.11<R_{hh}<12.64.  \nn
\end{align}

\noindent{\bf Collider simulation:~}%
Next we perform a detailed Monte Carlo simulation to estimate 
the needed integrated luminosity for probing or excluding the FNNP band
at the 14 TeV LHC. As a concrete example, we examine 
the $hh \to b\bar{b}\gamma\gamma$ channel, which has been 
studied by the ATLAS collaboration~\cite{ATL-PHYS-PUB-2014-019}.
MadGraph5~\cite{Alwall:2014hca} is used to generate 
the signal events at the parton-level with CT14~\cite{Dulat:2015mca} and 
MSTW2008~\cite{Martin:2009iq} parton distribution function (PDF).
Following the ATLAS study~\cite{ATL-PHYS-PUB-2014-019}, the signal 
events must contain two $b$-tagged jets and two isolated photons 
which satisfy the kinematic cuts as follows:
\bea
&&p_{T}^{{\rm leading}~b}>40~{\rm{GeV}},~
p_{T}^{b}>25~{\rm{GeV}},~|\eta^{b}|<2.5,\nonumber\\
&&p_{T}^\gamma>30~{\rm GeV},~|\eta^{\gamma}|<1.37~{\rm or}~
1.52<|\eta^{\gamma}|<2.37,\nonumber\\
&&\Delta R_0 <\Delta R_{b\bar b,\gamma\gamma}<2.0~, ~
\Delta R_{b\gamma}>\Delta R_0,~~ \Delta R_0=0.4~,\nonumber\\
&&100{\text{GeV}}<m_{b\bar b}<150{\text{GeV}},~
p_{T}^{b\bar b}>110{\text{GeV}},\nonumber\\
&&123{\text{GeV}}<m_{\gamma\gamma}<128{\text{GeV}},~
p_{T}^{\gamma\gamma}>110{\text{GeV}}.\nonumber
\label{eq:cuts}
\eea
In order to mimic the imperfect detector effects, we smear the final state parton momenta by a Gaussian distribution as suggested in Ref.~\cite{ATL-PHYS-PUB-2013-004}. The $b$ identification strongly depends on $p_T^{b}$ and $\eta^b$. We fit the $b$-tagging efficiency given in the ATLAS Technique Report~\cite{ATL-PHYS-PUB-2013-009} and obtain the following function of the $b$-tagging efficiency: 
\bea
\epsilon_b\left(p_{T},\eta\right)&=&0.135\tanh\left(\frac{p_{T}+50}
{75}\right)\tanh\left(\frac{450}
{p_{T}+80}\right)\nn\\
&\times& \big[3+e^{-\left(|\eta|-\sqrt{p_{T}/
1000}\right)^2/1.6}\big] e^{-|\eta|^3 p_{T}/1000}, ~~
\label{eq:btag}
\eea
where $p_T$ is in the unit of GeV. 
Figure~\ref{fig:cutacc} displays our $b$-tagging efficiency as a function of $p_T^b$ and $\eta^b$, which agrees well with the ATLAS study. 

\begin{figure}[h!]
\includegraphics[scale=0.2,clip]{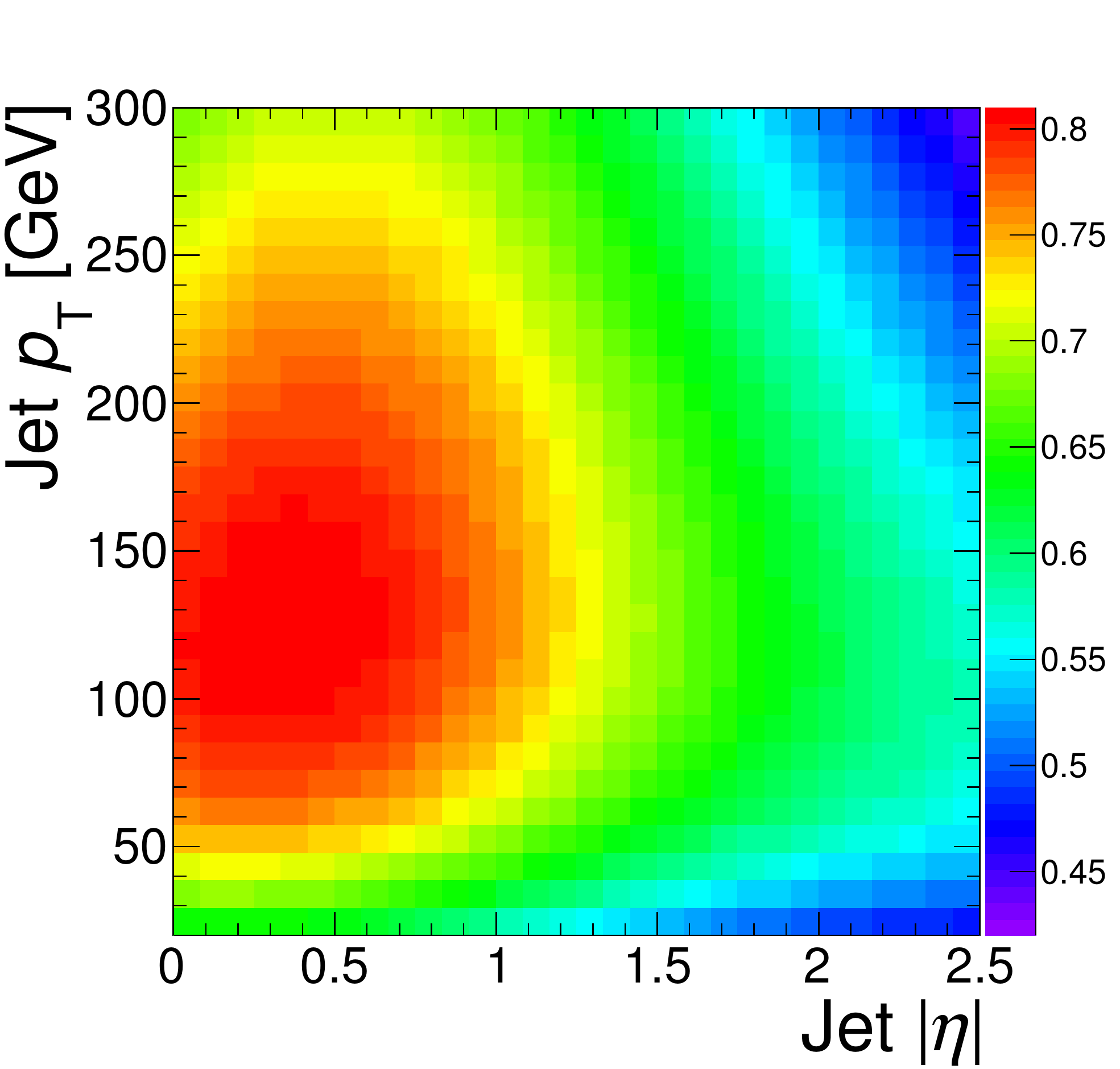}
\caption{The $b$-tagging efficiency as a function of $p_T^b$ and $\eta^b$.
\label{fig:cutacc} }
\end{figure}

The photon energy resolution and identification are crucial to trigger and reconstruct the signal events. In the simulation we adapt the photon energy resolution and identification efficiency given by the ATLAS collaboration~\cite{ATL-PHYS-PUB-2013-009}.  The photon energy resolution is given as follows:
 \bea
\sigma\left({\text{GeV}}\right)&=&0.3\oplus0.10\times\sqrt{E({\text{GeV}})}
\oplus0.010\times E({\text{GeV}}),\nonumber\\
&&~~{\text{for}}~~|\eta|<1.37,\nonumber\\
\sigma\left({\text{GeV}}\right)&=&0.3\oplus0.15\times\sqrt{E({\text{GeV}})}
\oplus0.015\times E({\text{GeV}}),\nonumber\\
&&~~{\text{for}}~~1.52<|\eta|<2.37.
\eea
The identification efficiency, which is sensitive to $p_T^\gamma$, is given by
\beq
\epsilon_\gamma\left(p_{T}\right)=0.76-1.98\exp\left(-\frac{p_{T}}
{16.1{\text{GeV}}}\right),
\eeq
Note that the identification rate is less than 80\% even for an energetic photon.

The key of collider simulation is to know the so-called cut efficiency, 
i.e. the fraction of signal events passing the kinematic cuts. 
To understand the cut efficiencies of different values of anomalous couplings,  one has to repeat the 
collider simulation which include all the kinematic cuts, imperfect detector resolutions and particle identifications, etc.  
However, it is very time consuming in practice. Inspired by the scalar feature of Higgs boson, we propose an analytical function to describe the fraction of signal events passing through the kinematic cuts. The function depends upon the invariant mass of the Higgs boson pair ($m_{hh}$) and is not sensitive to those anomalous couplings or specific NP model. The advantage of our method is 
that the cut efficiency of the $hh$ signal events can be easily obtained from the convolution of the differential cross section of $d\sigma/d m_{hh}$ and the cut efficiency function.  The method is explained below.

The scattering of $gg\to hh$ is dominated by the 
$s$-wave contribution. Owing to the scalar feature of the Higgs boson, 
there is no spin correlations among the initial state and final state particles. 
Therefore, the $p_{\rm T}$ and $\eta$ distributions of the $b$-jets and 
photons depend mainly upon $m_{hh}$. The differential 
cross section of the $gg\to hh$ process before any cut is
\bea
\frac{d\sigma}{d\mhh}&=&\frac{m_{hh}}{S^2}{\cal H}\left(\mhh
,\mu_r\right)\int^1_{\mhh^2/S} \frac{dx_1}{x_1}
f_{g/p}\left(\frac{m_{hh}^2}{x_1 S},\mu_f\right)
\nonumber\\&&
\times f_{g/p}\left(x_1,\mu_f\right)
\int d\eta \left|\frac{\partial\hat\eta}
{\partial\eta}\right|_{m_{hh},\eta,x_1}\nonumber\\
&\equiv&\frac{m_{hh}}{S^2}{\cal H}\left(m_{hh}
,\mu_r\right)\Sigma\left(m_{hh},S,\mu_f\right).
\eea
where $\eta$ and $\hat{\eta}$ is the rapidity of one of the Higgs bosons 
in the laboratory  frame and center of mass (c.m.) frame, respectively. $\sqrt{S}$ is the collision 
energy of the hadron collider, ${\cal H}(m_{hh}, \mu_r)$ is the hard scattering cross section with $\mu_r$ the renormalization scale,  $f_{g/p}$ is the gluon PDF with $\mu_f$ the factorization scale.
As argued above, the cut efficiency depends on the 
configuration of the Higgs bosons which is described 
by $\mhh$ and $\eta$, and the differential cross section after cuts is
\bea
\frac{d\sigma_{\text{cut}}}{dm_{hh}}&=&\int d\tilde{m}_{hh}
\frac{\tilde{m}_{hh}}
{S^2}{\cal H}\left(\tilde{m}_{hh}
,\mu_r\right)\int^1_{\tilde{m}_{hh}^2/S} \frac{dx_1}{x_1}
\nonumber\\&&
\times f_{g/p}\left(\frac{\tilde{m}_{hh}^2}{x_1 S},\mu_f\right)f_{g/p}\left(x_1,\mu_f\right)
\int d\eta \nonumber\\&&
\times\left|\frac{\partial\hat\eta}
{\partial\eta}\right|_{\tilde{m}_{hh},\eta,x_1}
\epsilon\left(m_{hh},\tilde{m}_{hh},x_1,\eta\right),
\eea
where $\tilde m_{hh}$ is introduced to describe the finite energy smearing effect and $\epsilon$ stands for the cut acceptance. Define 
\bea
\tilde\Sigma\left(\tilde{m}_{hh},S,\mu_f\right)&\equiv&
\int dm_{hh}\int^1_{\tilde{m}_{hh}^2/S} \frac{dx_1}{x_1}
f_{g/p}\left(\frac{\tilde{m}_{hh}^2}{x_1 S},\mu_f\right)\nonumber\\&&
\times f_{g/p}\left(x_1,\mu_f\right)
\int d\eta \left|\frac{\partial\hat\eta}
{\partial\eta}\right|_{\tilde{m}_{hh},\eta,x_1}
\nonumber\\&&
\times \epsilon\left(m_{hh},\tilde{m}_{hh},x_1,\eta\right),
\eea
then 
\be
\sigma_{\text{cut}}=\int d\tilde{m}_{hh}\frac{\tilde{m}_{hh}}{S^2}
{\cal H}\left(\tilde{m}_{hh}
,\mu_r\right)\tilde\Sigma\left(\tilde{m}_{hh},S,\mu_f\right).
\ee
We introduce a differential cut efficiency function 
\be
{\mathcal{A}}\left(m_{hh},S,\mu_f\right)
=\frac{\tilde\Sigma\left(m_{hh},S,\mu_f\right)}
{\Sigma\left(m_{hh},S,\mu_f\right)}\equiv \mathcal{A}(m_{hh}),
\label{eq:defA}
\ee
which depends on $\sqrt{S}$, parton distribution functions and kinematic cuts. 
Again, we emphasize that the $\mathcal{A}$-function is model 
independent. If the $gg\to hh$ scattering in NP models is also dominated by the 
$s$-wave scattering, its production rate after cuts is
\bea
\sigma_{\text{cut}}=\int dm_{hh}\frac{d\sigma}{dm_{hh}}\otimes
{\mathcal{A}}\left(m_{hh},S,\mu_f\right).
\label{eq:convolution}
\eea
\begin{figure*}
\includegraphics[scale=0.23,clip]{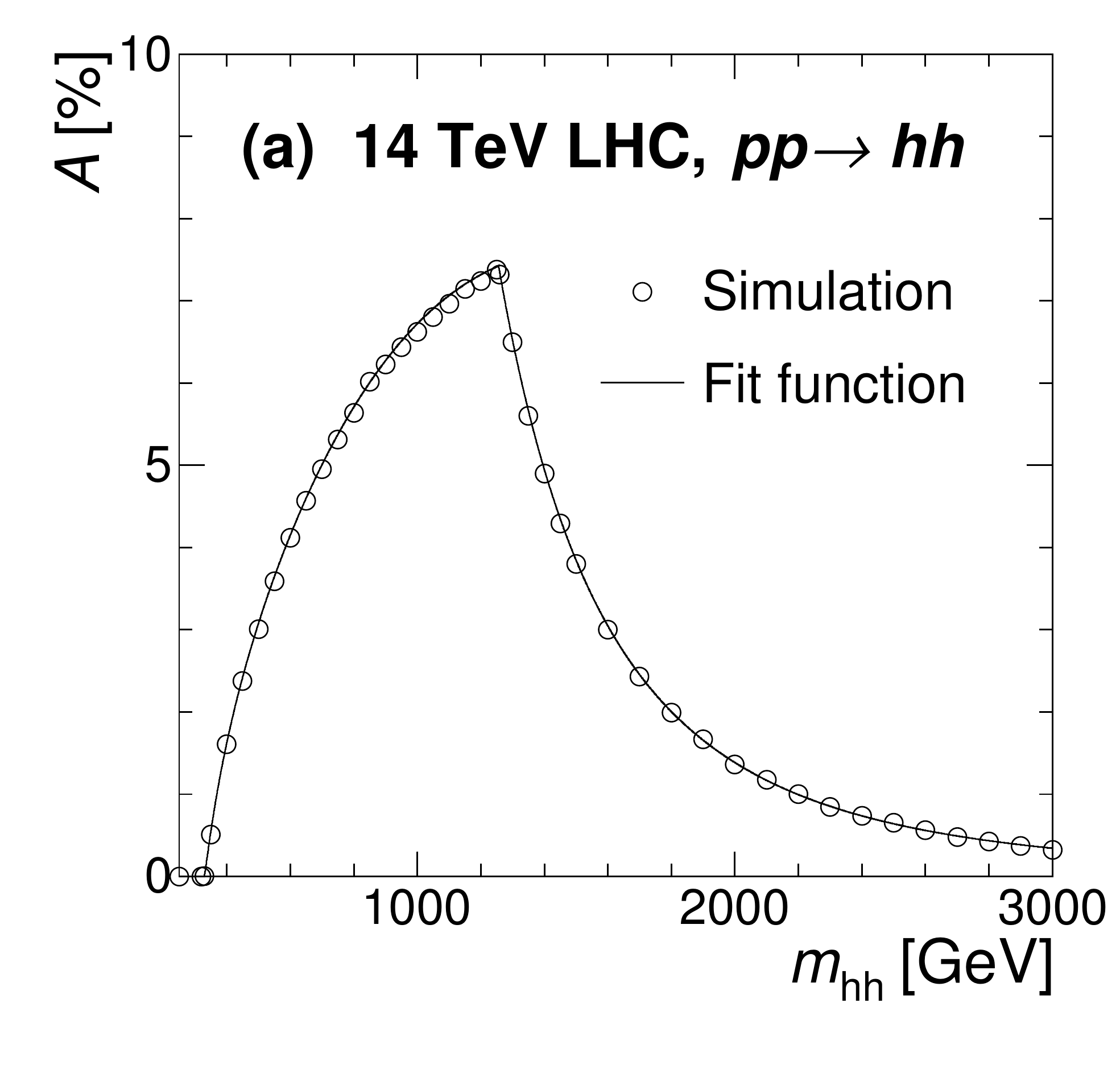}
\includegraphics[scale=0.23,clip]{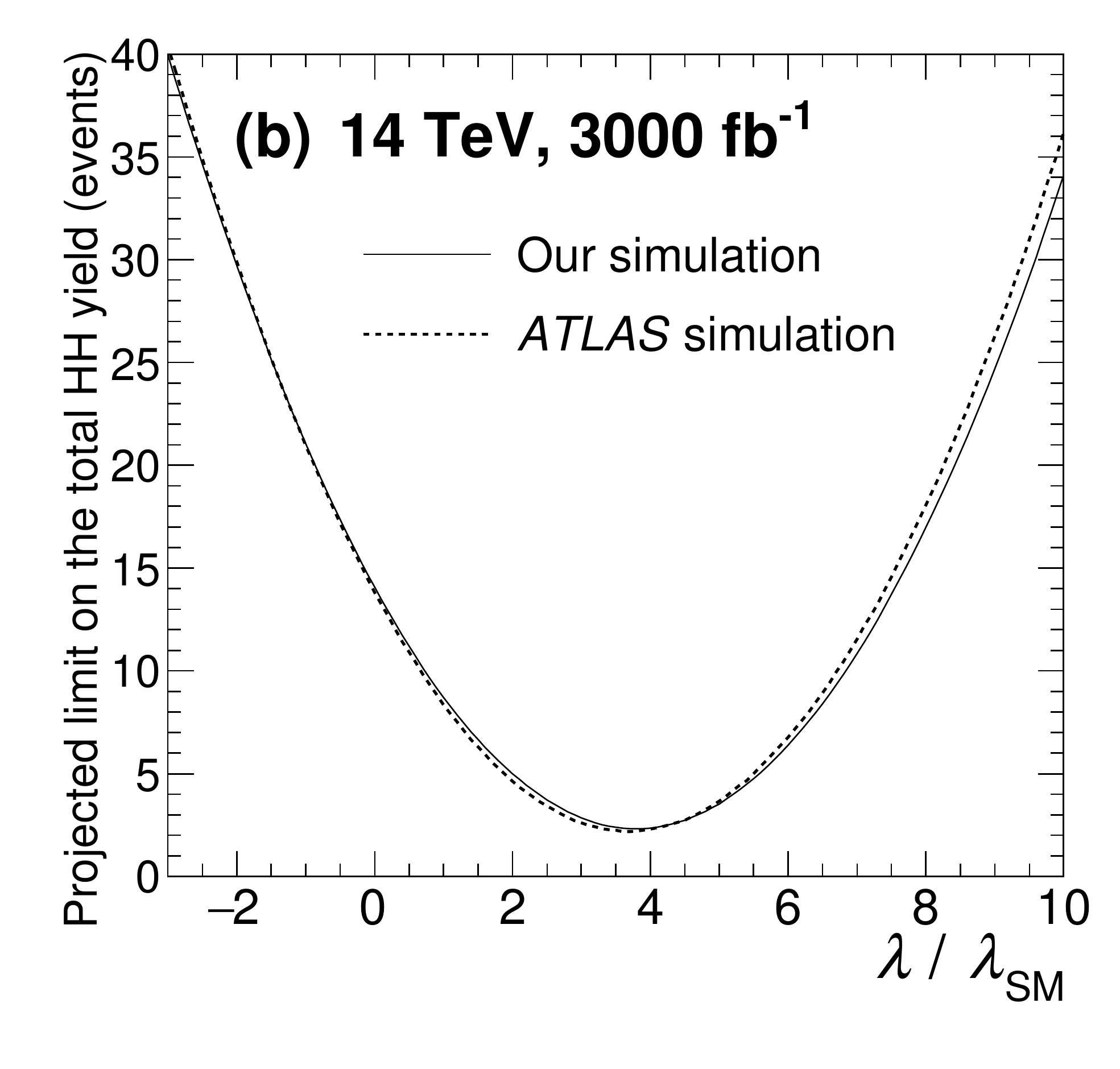}
\includegraphics[scale=0.235,clip]{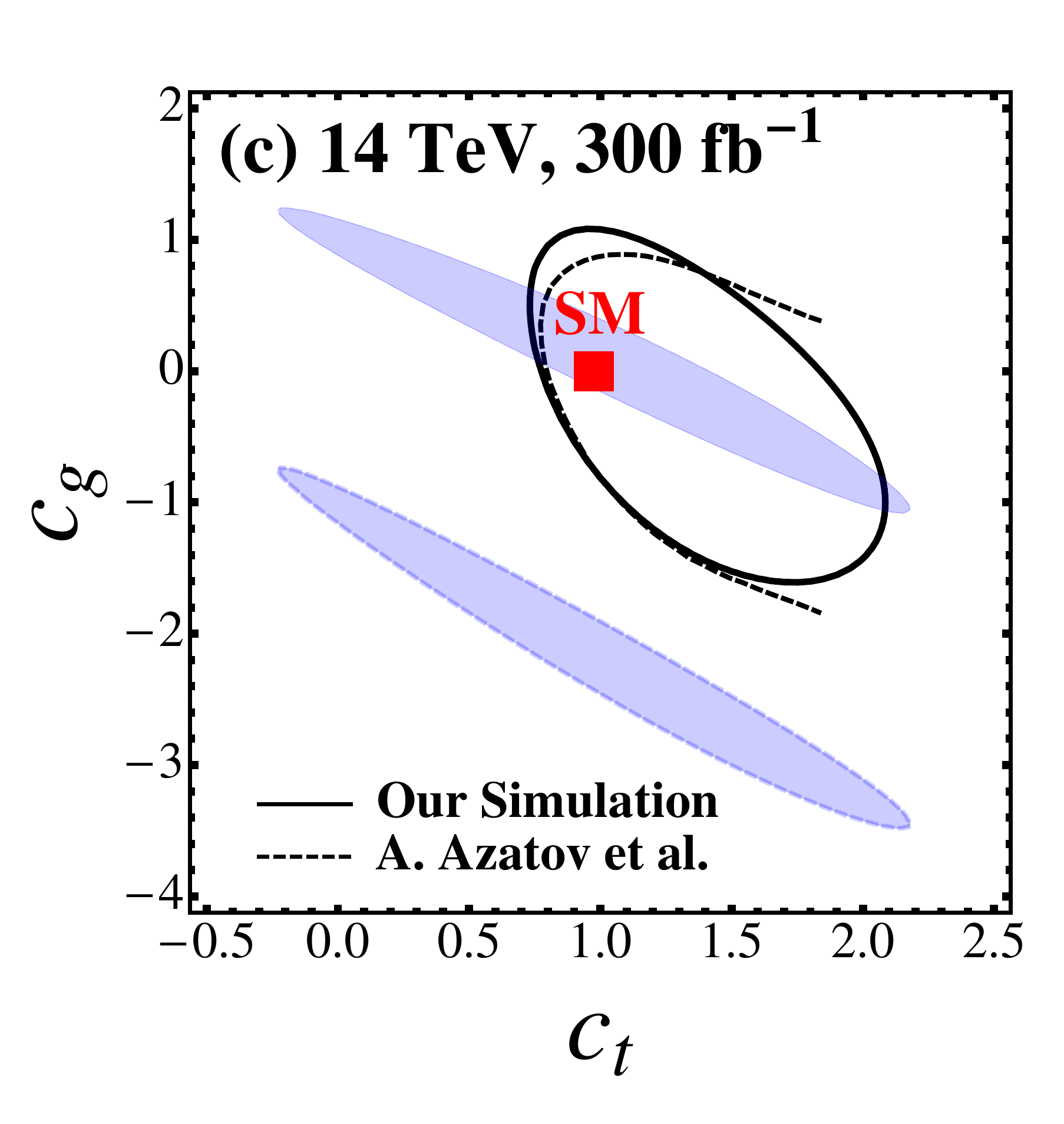}
\caption{(a): Cut efficiency as a function 
of $m_{hh}$ at the 14TeV LHC; (b): comparison between our 
method (solid curve) and the ATLAS simulation (dashed 
line, see Fig. 8 in \cite{ATL-PHYS-PUB-2014-019}); (c) comparison 
of 68\% C.L. exclusion contours in the plane of $c_g$ and $c_t$ 
obtained in our method (solid curve) and the one in 
\cite{Azatov:2015oxa} (dashed curve) for the choice 
$(b)$, i.e. $c_{2h}=3(c_t-1)/2$.
\label{fig:cutacc} }
\end{figure*}%
The $\mathcal{A}$-function could be derived analytically for a given $m_{hh}$. It is hard to account for the detector effects analytically, however. In this work we first obtain the cut efficiency of signal events from the Monte Carlo simulation with all the detector effects and then fit the efficiency with the following functions as suggested by analytical calculations:
\begin{widetext}
\beq
\mathcal{A}\left(m_{hh}\right)=
\begin{cases}
c_1\left[1-\sqrt{\dfrac{m_{hh}^2\left(1-\cos\Delta R_0\right)
-8\left(m_h-\delta m_1\right)^2}{\left(1-\cos\Delta R_0\right)\left(m_{hh}^2
-4\left(m_h-\delta m_1\right)^2\right)}}\right]^{\gamma_c}, & m_{hh}>m^{(t)}_{hh},\\
c_2 \left[1-\displaystyle \frac{4p_{{\rm T},h}^2}{m_{hh}^2-4\left(m_h
-\delta m_2\right)^2}\right]^{\beta_a}\left(\dfrac{m_{hh}}{\sqrt{S}}\right)^{\beta_b}
\left[1+\beta_c\left(\dfrac{m_{hh}}{\sqrt{S}}\right)\log
\left(\dfrac{2m_{hh}}{\sqrt{S}}\right)\right],& 329.3{\text{GeV}}<m_{hh}<m^{(t)}_{hh},\\
0,& m_{hh}<329.3{\text{GeV}}.
\end{cases}
\label{eq:cuteff}
\eeq
\end{widetext}
The $\gamma_c$, $\beta_{a,b,c}$ and $\delta m_{1,2}$ parameters 
reflect the imperfect detector effects, $c_{1,2}$ are the normalization parameters and
$m^{(t)}_{hh}$ is the turning point of two fitting functions.

\begin{table}[b]
\caption{Fitting parameters (the second and fifth rows) of $\mathcal{A}(m_{hh})$ and the corresponding uncertainties (the third and sixth rows) for two PDF sets: (top) CT14~\cite{Dulat:2015mca} and (bottom) MSTW2008~\cite{Martin:2009iq}.}
\begin{center}
\begin{tabular}{c|ccc|ccccc}
\hline
~~$m_{hh,0}$~~&~~$c_1$~~&~~$m_1$~~&~~$ \gamma_c$~~&
~~$c_2$~~&~~$\delta m_2$~~
&~$\beta_a$~&~$\beta_b$~&~$\beta_c$~\\
\hline
1.257TeV & 1.1378 & 50GeV & 1.675 & 11.02 & 2.5GeV & 1.13 & 1.48 & 4.88 \\
\hline
$-$ & 0.0037 & $-$ & 0.002 & 1.20 & $-$ & 0.02 & 0.02 & 0.03 \\
\hline
\end{tabular}\vspace*{0.5mm}
\begin{tabular}{c|ccc|ccccc}
\hline
~~$m_{hh,0}$~~&~~$c_1$~~&~~$m_1$~~&~~$ \gamma_c$~~&
~~$c_2$~~&~~$\delta m_2$~~
&~$\beta_a$~&~$\beta_b$~&~$\beta_c$~\\
\hline
1.254TeV & 1.1419 & 50GeV & 1.677 & 9.66 & 2.5GeV & 1.13 & 1.44 & 4.82 \\
\hline
$-$ &0.0040 & $-$ & 0.002 & 1.09 & $-$ & 0.02 & 0.02 & 0.03 \\
\hline
\end{tabular}
\end{center}
\label{tbl:fitting}
\end{table}

Figure~\ref{fig:cutacc}(a) displays the cut efficiency as a function of $m_{hh}$ from the Monte Carlo simulation result (circle).   
We note that the efficiency depends mainly on the cut on $p_T^{b\bar{b},\gamma\gamma}$ and $\Delta R$.  The large $p_T^{b\bar{b},\gamma\gamma}$ cuts require the Higgs boson's energy  must be more than $\sqrt{m_h^2+p_{T}^2}\sim 167$ GeV such that $m_{hh}>330~{\rm GeV}$. Therefore, the cut efficiency is zero for $m_{hh}<330~{\rm GeV}$. The efficiency increases with $m_{hh}$ because  the $p_T$'s of Higgs boson decay products also increase with $m_{hh}$ such that more signal events pass the $p_T$ cuts. On the other hand, $\Delta R_{b\bar{b}}$ decreases with $m_{hh}$ and reaches  $\Delta R_0 = 0.4$ at $m_{hh}=m^{(t)}_{hh}$. For $m_{hh}>m^{(t)}_{hh}$ the cut efficiency decreases with $m_{hh}$ because $\Delta R_{b\bar{b}}$ is likely to be smaller than 0.4 such that most of the signal events fail the $\Delta R$ cut. We fit the Monte Carlo data with the $\mathcal{A}$-function in Eq.~\ref{eq:cuteff} and obtain those fitting parameters which are shown in Table~\ref{tbl:fitting}. We note that the fitting parameters are not sensitive to the PDF sets. Both CT14 and MSTW2008 PDF sets yield similar parameters. For comparison we also plot the fitting function in Fig.~\ref{fig:cutacc}(a);  see the solid curve.

In order to check our method, we compare our results with those of 
the ATLAS collaboration~\cite{ATL-PHYS-PUB-2014-019} and Ref.~\cite{Azatov:2015oxa}. The comparisons are shown in Fig.~\ref{fig:cutacc}(b) and Fig.~\ref{fig:cutacc}(c), respectively. 
Our results are consistent with those results in~\cite{ATL-PHYS-PUB-2014-019,Azatov:2015oxa}.
Figure~\ref{fig:cutacc}(c) shows the exclusion contours at 68\% C.L. 
with $\mathcal{L}=300~{\rm fb}^{-1}$ for the case $(b)$ of anomalous couplings.

\noindent{\bf Conclusion and Discussion:}~%
Now equipped with the cut efficiency function $\mathcal{A}(m_{hh})$, we are ready to estimate the potential of excluding the FNNP band at the LHC.  The SM backgrounds include $b\bar{b}\gamma\gamma$, $c\bar{c}\gamma\gamma$, $b\bar{b}\gamma j$, $jj\gamma\gamma$, $b\bar{b}jj$, $t\bar{t}(\geqslant 1\ell^\pm)$, $t\bar{t}\gamma$, $Z(\to b\bar{b})h(\to \gamma\gamma)$, $t\bar{t}h(\to \gamma\gamma)$ and $b\bar{b}h(\to \gamma\gamma)$, etc.
There are 4.72 background events after all the cuts at the 14~TeV LHC with an integrated luminosity of $300~{\rm fb}^{-1}$~\cite{ATL-PHYS-PUB-2014-019}. 
We calculate the 95\% C.L. exclusion bound with
\beq
\sqrt{-2\left(n_b\ln\dfrac{n_s+n_b}{n_b}-n_s\right)}=1.96,
\eeq
where $n_s$ and $n_b$ denotes the number of signal and background events, respectively.

\begin{figure}
\includegraphics[scale=0.24,clip]{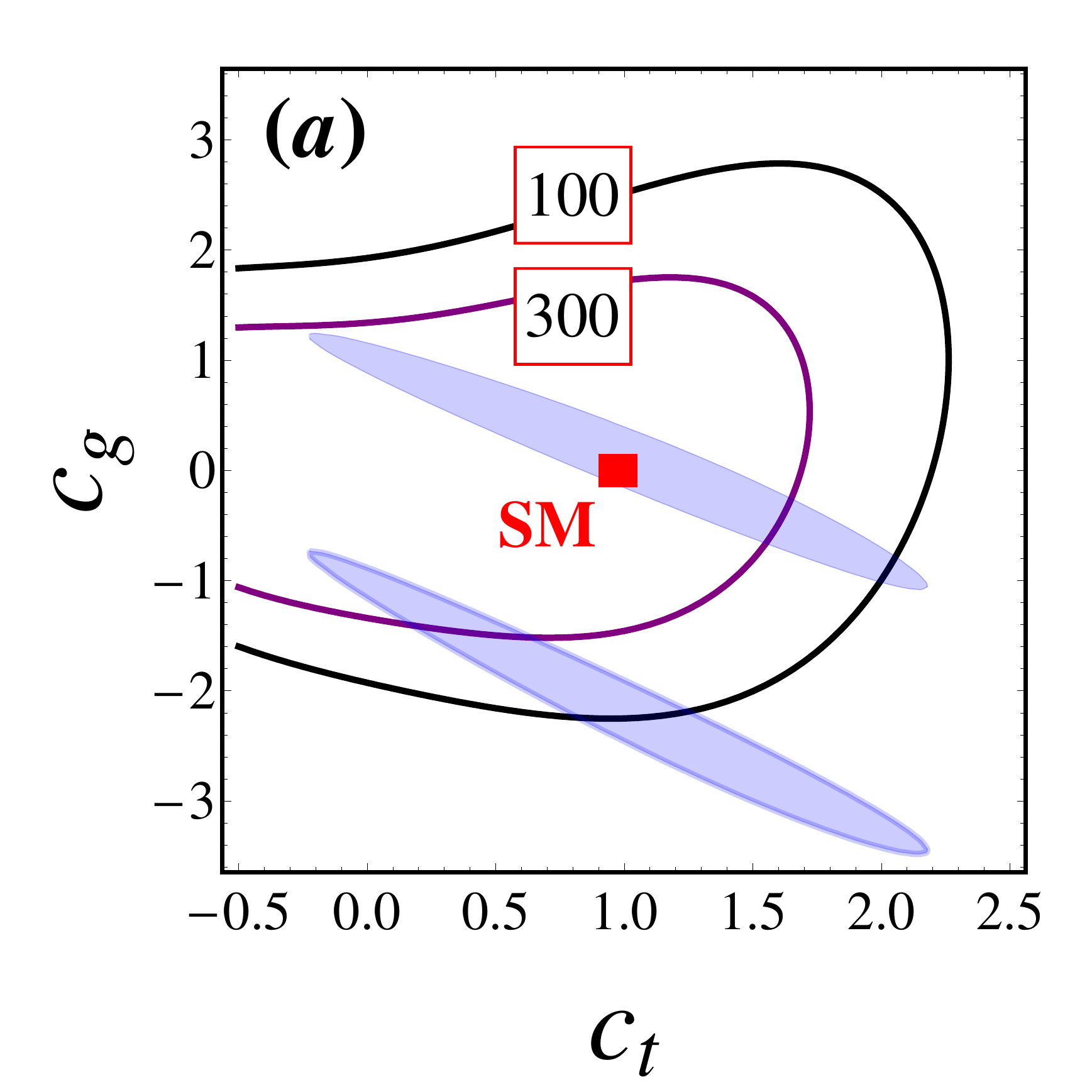}
\includegraphics[scale=0.24,clip]{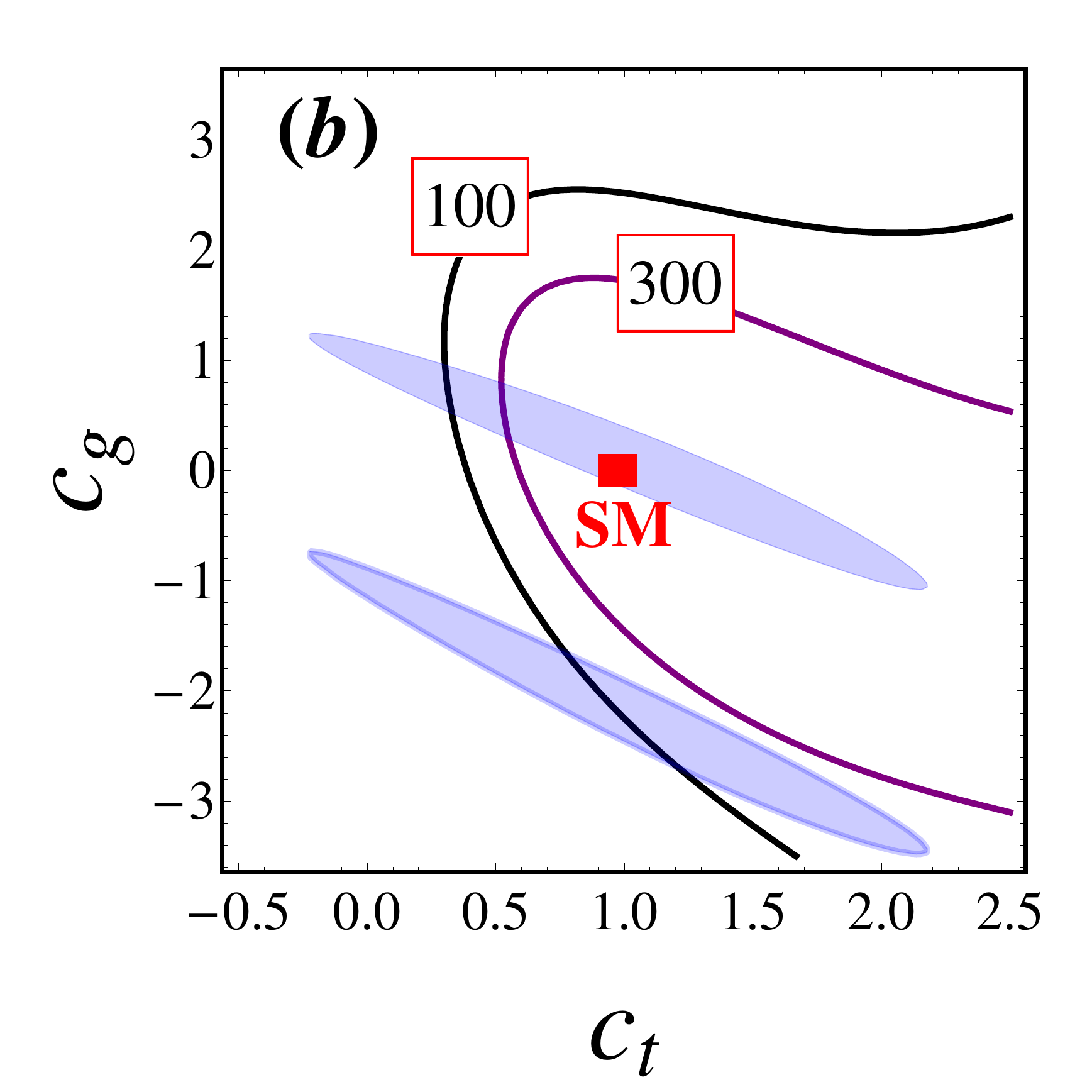}
\includegraphics[scale=0.24,clip]{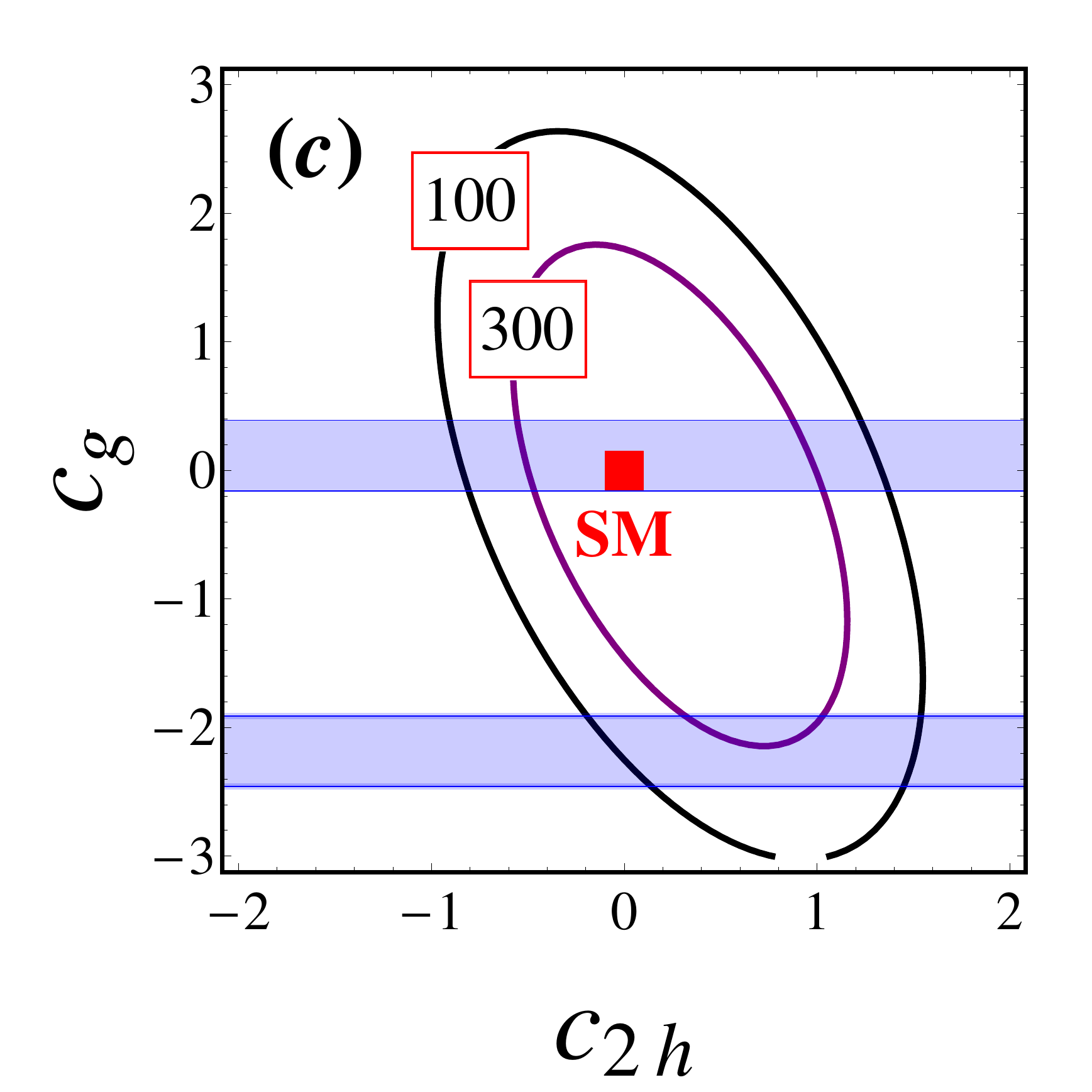}
\includegraphics[scale=0.24,clip]{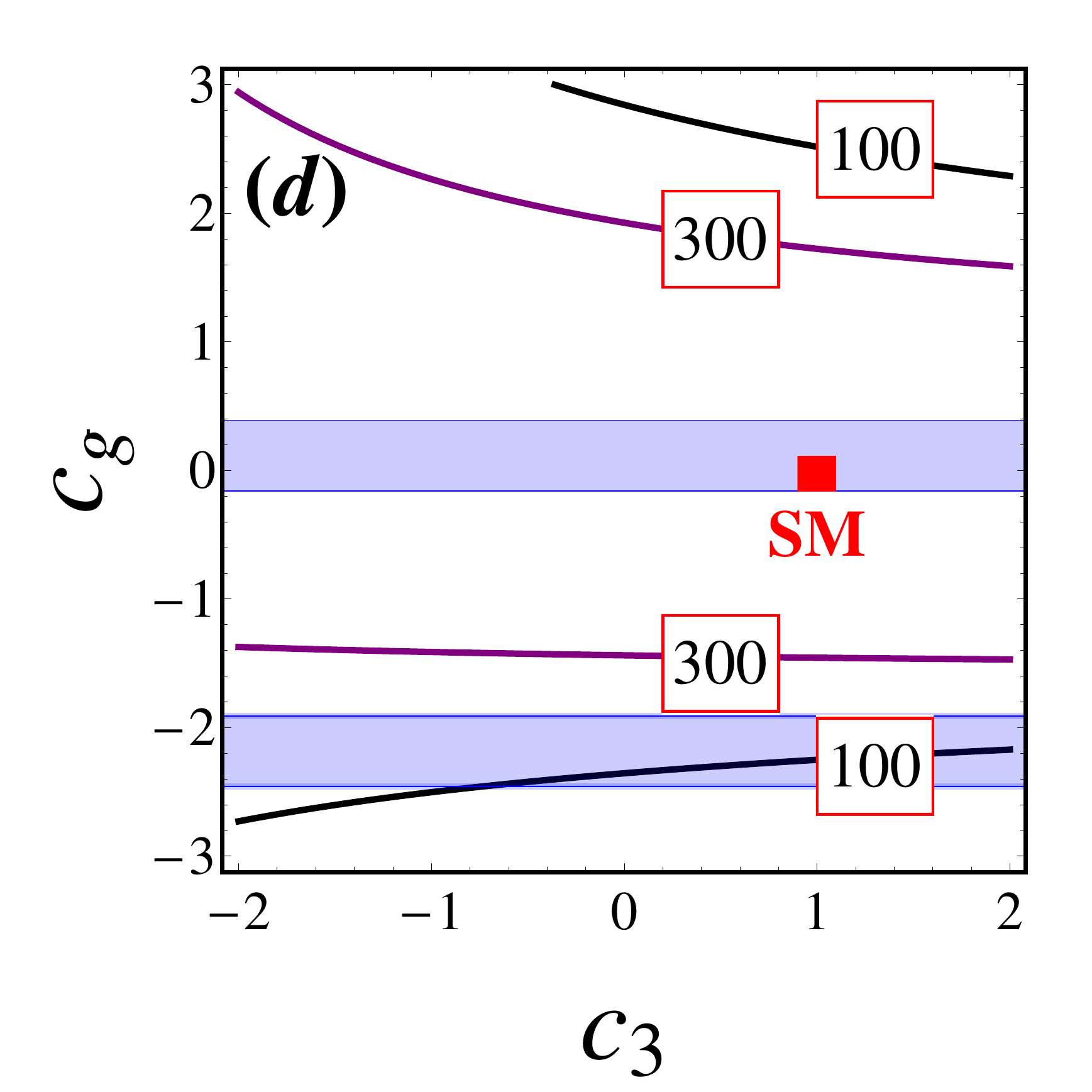}
\caption{The 95\% C.L. exclusion contours in the plane of anomalous couplings obtained from the  
Higgs boson pair production with luminosity $\mathcal{L}=$ 100 fb$^{-1}$ 
and 300 fb$^{-1}$ at the 14 TeV LHC. The figure indices correspond to the benchmark choices of anomalous couplings in Eq.~\ref{eq:benchmark}.}
\label{fig:L14} 
\end{figure}

Figure~\ref{fig:L14} displays the contours of 95\% C.L. exclusion bound 
from the Higgs boson pair production with $\mathcal{L}=100~{\rm fb}^{-1}$ and $300~{\rm fb}^{-1}$
at the 14 TeV LHC for the four choices of anomalous couplings. 
The exclusion contours of the cases $(a)$ and $(b)$ are different from those $R_{hh}$ contours shown in Fig.~\ref{fig:R14}. 
The differences occur in the large positive $c_t$ region where ${\rm BR}(h\to \gamma\gamma)$ is highly reduced. A large $c_g$ is needed to reach the same rate of the Higgs boson pair production.

In the case $(c)$ the shapes of the exclusion bounds are very similar to those of the $R_{hh}$ contours; see Fig.~\ref{fig:L14}(c) and Fig.~\ref{fig:R14}(c). It can be understood from the fact that the $c_{2h}$ is not sensitive to the kinematic cuts or Br$(h\to \gamma\gamma)$. 

In the case $(d)$, the shape of exclusion bounds in Fig.~\ref{fig:L14}(d) is slightly different from that of the $R_{hh}$ contours in Fig.~\ref{fig:R14}(d), especally in the small $c_3$ region.  The difference can be understood from the cuts we imposed. The $c_3$ coupling contributes sizeably in the small $m_{hh}$ region. However, the hard $p_T$ cut on $b\bar{b}$ and $\gamma\gamma$ pairs demands a large $m_{hh}$ region. As a result, the exclusion contours depend mildly on $c_3$. It is consistent with Ref.~\cite{Contino:2012xk}.

If no excess were observed in the Higgs boson pair production, then one can impose constraints on the anomalous couplings, especially on the FNNP band. The minimal luminosities to exclude the FNNP band at 95\% C.L. in the four choices of anomalous couplings are 
\begin{align}
& (a):~ \mathcal{L}\geq 1681~{\rm fb}^{-1}; && (b): ~\mathcal{L}\geq 235~{\rm fb}^{-1}; \nn\\
& (c): ~\mathcal{L}\geq 446~{\rm fb}^{-1};   && (d): ~\mathcal{L}\geq 186~{\rm fb}^{-1}.\nn
\end{align}
It is worth mentioning that the current allowed bands shown in Fig.~\ref{fig:globalfit}
would shrink when the LHC Run-2 data comes out.  It is possible to exclude the FNNP band
with a luminosity smaller than the values shown above.

\vspace*{3mm}
\noindent {\it Note:}~While this article was in finalization, 
the paper by B. Batell {\it et al}~\cite{Batell:2015koa} appeared online and
investigated the correlation between single Higgs and double Higgs 
boson production in supersymmetric model.

\vspace*{3mm}
\noindent{\bf Acknowledgments:}~%
The work of QHC, BY and DMZ is supported in part by the National 
Science Foundation of China under Grand No. 11275009. HZ is supported 
by the U.S. DOE under contract No. DE-SC0011702. 
Part of this work was done while HZ was visiting the Center for High Energy Physics at Peking University. 
HZ is pleased to recognize this support and the hospitality of the Center of HEP at Peking University.

\bibliographystyle{apsrev}
\bibliography{reference}

\end{document}